\documentclass{Interspeech}

\interspeechcameraready

\title{Universal Speech Enhancement with Regression and Generative Mamba}

\author[affiliation={1, 2}]{Rong}{Chao}
\author[affiliation={3}]{Rauf}{Nasretdinov}
\author[affiliation={3}]{Yu-Chiang Frank}{Wang}
\author[affiliation={3}]{Ante}{Jukić}
\author[affiliation={3}]{Szu-Wei}{Fu}
\author[affiliation={1}]{Yu}{Tsao}

\affiliation{CITI}{Academia Sinica}{Taiwan}
\affiliation{CSIE}{National Taiwan University}{Taiwan}
\affiliation{}{NVIDIA}{}

\email{d13922037@ntu.edu.tw, yu.tsao@citi.sinica.edu.tw}

\usepackage{comment}
\usepackage{amsmath,graphicx}
\usepackage{booktabs}
\usepackage{graphicx}
\usepackage{xcolor}
\usepackage{amsmath, amssymb, graphicx}
\usepackage{algorithm}
\usepackage{algpseudocode}
\usepackage{url}
\usepackage{ragged2e}
\usepackage{balance}
\usepackage{multirow}
\usepackage{amssymb}
\usepackage{subfig}
\usepackage{makecell} 

\usepackage{booktabs}

\begin{document}
%
\setlength{\lightrulewidth}{0.02em}
\setlength{\heavyrulewidth}{0.08em}
%
\setlength{\textfloatsep}{5.0pt plus 3.0pt minus 3.0pt}
\setlength{\floatsep}{5.0pt plus 3.0pt minus 3.0pt}
\setlength{\intextsep}{3.0pt plus 3.0pt minus 3.0pt}
\setlength{\dbltextfloatsep}{6.0pt plus 3.0pt minus 2.0pt}
\maketitle

\begin{abstract}
The Interspeech 2025 URGENT Challenge aimed to advance universal, robust, and generalizable speech enhancement by unifying speech enhancement tasks across a wide variety of conditions, including seven different distortion types and five languages. We present Universal Speech Enhancement Mamba (USEMamba), a state-space speech enhancement model designed to handle long-range sequence modeling, time-frequency structured processing, and sampling frequency-independent feature extraction. Our approach primarily relies on regression-based modeling, which performs well across most distortions. However, for packet loss and bandwidth extension, where missing content must be inferred, a generative variant of the proposed USEMamba proves more effective. Despite being trained on only a subset of the full training data, USEMamba achieved \textbf{2nd place in Track 1} during the blind test phase, demonstrating strong generalization across diverse conditions. 

    
    

\end{abstract}

\section{Introduction}
Speech enhancement (SE) aims to improve the intelligibility and perceptual quality of captured speech signals which may include a range of degradations. It is a fundamental task in speech processing, playing a crucial role in applications such as hearing aids \cite{HA}, automatic speech recognition (ASR) \cite{wang2014training}, and speaker verification \cite{SR}. Over the past few decades, SE has undergone notable advances, transitioning from early statistical signal processing-based methods to deep learning-driven approaches~\cite{gannot2024spm}. 



Recently, transformer-based SE models \cite{dang2022dpt, CMGAN} have demonstrated strong sequence modeling capabilities, using self-attention mechanisms to capture long-range dependencies. However, their quadratic computational complexity limits the scalability of audio length. To address this, state-space models (SSMs) \cite{gu2023mamba} have emerged as an alternative, offering efficient long-range sequence modeling while maintaining a lower computational overhead. The Structured State Space Model (S4) \cite{gu2022parameterization} and its recent extension, Mamba \cite{gu2023mamba}, have shown promising results in speech and audio processing tasks \cite{ku2023multi, du2024spiking, sun2024dual}. These advancements highlight the shift toward models that balance computational efficiency and robust speech enhancement.

Despite these advancements, achieving universality, robustness, and generalizability in SE remains a critical challenge. Most SE models are trained on datasets with a single acoustic condition, a specific type of distortion, and a single language, which constrains their generalization to diverse real-world scenarios \cite{wang2018supervised}. Addressing these challenges requires models that can adapt to different acoustic conditions, sampling configurations, and languages.

To push the limits of SE generalization, the URGENT 2025 Challenge was first introduced at Interspeech 2024 \cite{zhang2024urgent}. The competition evaluated SE models under seven distinct distortions (additive noise, reverberation, clipping, bandwidth limitation, codec artifacts, packet loss, and wind noise) across five languages (English, German, French, Spanish, and Chinese) with diverse sampling frequencies (8, 16, 22.05, 24, 32, 44.1, and 48 kHz). 

In this work, we propose Universal Speech Enhancement Mamba (USEMamba), a state-space model-based speech enhancement system built upon the recently developed Mamba architecture \cite{gu2023mamba}. Unlike conventional Transformer-based models, Mamba facilitates efficient long-range sequence modeling while notably reducing computational cost. Therefore, a deeper network architecture can be adopted to enhance the modeling capacity, allowing for more effective speech enhancement. USEMamba incorporates selective state-space modeling to enable efficient speech signal processing and leverages time-frequency Mamba blocks to jointly capture spectral and temporal dependencies. To improve generalization across diverse conditions, the model employs a sampling frequency-independent (SFI) feature extraction mechanism, ensuring robustness across varying sampling rates. 

Our participation in the URGENT 2025 Challenge demonstrated the effectiveness of USEMamba, where it achieved \textbf{2nd place in both non-blind and blind test phases}, and 5th place in the final ranking when the subjective SE metric was included, highlighting its strong generalization ability and robustness. 


\begin{figure*}[t]
    \centering
    \includegraphics[width=0.95\textwidth]{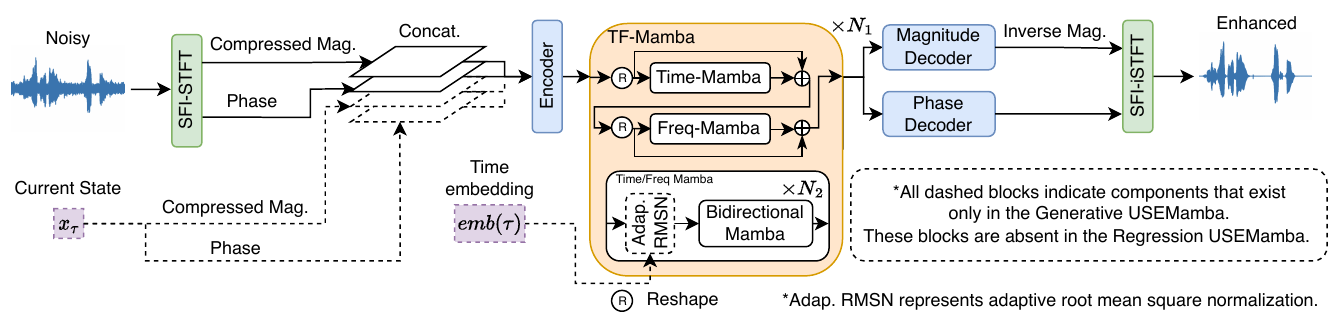}
    \caption{Overview of the Universal SEMamba architecture.}
    \label{fig:semamba_architecture}
\end{figure*}

\section{Related Work}
In recent years, researchers have proposed joint frameworks that integrate multiple enhancement tasks into a single model to improve generalization ability \cite{serra2022universal, koizumi2023miipher, su2020hifi, liu2022voicefixer, li2024masksr, zhang2025anyenhance, zhang2023toward, wang2024speechx, yanguniaudio, ku2024generative}. HiFi-GAN \cite{su2020hifi} employs adversarial networks for denoising and dereverberation. VoiceFixer \cite{liu2022voicefixer} utilizes a ResUNet-based analysis stage and a neural vocoder-based synthesis stage to address various speech enhancement tasks, including denoising, dereverberation, declipping, and bandwidth extension. Similarly, MaskSR \cite{li2024masksr} and AnyEnhance \cite{zhang2025anyenhance} leverage masked generative modeling for full-band speech restoration, with the latter also incorporating target speaker extraction. Zhang et al. \cite{zhang2023toward} introduced a universal SE model designed for both denoising and dereverberation while accommodating diverse input formats (e.g., multi-channel, multi-sampling rate). Additionally, Ku et al. \cite{ku2024generative} proposed a flow matching-based model pretrained on a large-scale dataset, which can be fine-tuned for various downstream restoration tasks.

\subsection{SEMamba: Incorporating Mamba in Speech Enhancement}

Recent advancements in SSMs have led to their adoption in speech processing tasks, offering an efficient alternative to Transformer-based architectures. One such model, SEMamba-advanced (denoted for brevity as SEMamba in this manuscript), introduced the Mamba structured SSM into SE \cite{chao2024investigation}. By leveraging long-range sequence modeling with linear-time complexity, SEMamba notably reduces computational overhead compared to self-attention-based SE approaches~\cite{chao2024investigation}.


The core of SEMamba is formulated using a SSM, where the latent state $\mathbf{h}_t$ is updated via:
\begin{equation}
    \mathbf{h}_{t+1} = \bar{A} \mathbf{h}_t + \bar{B} \mathbf{x}_t,
\end{equation}
\begin{equation}
    \mathbf{y}_t = C \mathbf{h}_t,
\end{equation}
where $\bar{A}$ and $\bar{B}$ are discretized state matrices, and $C$ is the output transformation matrix \cite{gu2023mamba}. Unlike Transformers, which rely on self-attention, SEMamba efficiently propagates sequence information through selective gating mechanisms, reducing computational overhead while maintaining long-range dependencies.



\begin{table*}[h]
    \centering
    \caption{Dataset Composition for URGENT 2025 Challenge. We used only a subset of the full training data, as highlighted in bold. }
    \vspace{-0.6em}
    \resizebox{0.90\textwidth}{!}{%
    \begin{tabular}{lllcc}
        \toprule
        \textbf{Type} & \textbf{Corpus} & \textbf{Condition} & \textbf{Sampling (kHz)} & \textbf{Duration (h)} \\
        \midrule
        \multirow{6}{*}{Speech} 
        & \textbf{LibriVox (DNS5)} & Audiobook & 8--48 & 350 \\
        & \textbf{LibriTTS} & Audiobook & 8--24 & 200 \\
        & \textbf{VCTK} & Newspaper-style & 48 & 80 \\
        & \textbf{WSJ} & WSJ news & 16 & 85 \\
        & EARS & Studio recording & 48 & 100 \\
        & Multilingual Librispeech (de, en, es, fr) & Audiobook & 8--48 & 450 \\
        & \textbf{CommonVoice} 19.0 (de, \textbf{en}, es, fr, zh-CN) & Crowd-sourced voices & 8--48 & 1300 \\
        \hline
        \multirow{4}{*}{Noise} 
        & \textbf{AudioSet+FreeSound (DNS5)} & Crowd-sourced + YouTube & 8--48 & 180 \\
        & \textbf{WHAM! Noise} & 4 urban environments & 48 & 70 \\
        & FSD50K (Filtered) & Crowd-sourced & 8--48 & 100 \\
        & Free Music Archive & Directed by WFMU & 8--44.1 & 200 \\
        & Wind noise simulated by participants & - & Any & - \\
        \hline
        \multirow{2}{*}{RIR} 
        & \textbf{Simulated RIRs (DNS5)} & SLR28 & 48 & 60k samples \\
        & RIRs simulated by participants & - & Any & - \\
        \bottomrule
    \end{tabular}
    }
    \label{tab:dataset}
\end{table*}

\section{Universal SEMamba Architecture}

Building upon the SEMamba framework, we propose \textbf{Universal SEMamba (USEMamba)}, an improved model designed for universal speech restoration by introducing key architectural and training modifications. The enhancements focus on 1) refining loss functions, 2) transitioning from masking-based to mapping-based magnitude prediction, 3) incorporating sampling frequency-independent (SFI) feature extraction, 4) scaling up the model size by increasing the depth of the Time-Frequency Mamba (TF-Mamba) module, and 5) propose a generative variant called USEMamba-Flow. Figure~\ref{fig:semamba_architecture} shows the overall model framework.

\subsection{Loss Function Design}
The training objective of USEMamba is formulated as a combination of time domain $L_{1}$-norm loss, multi-resolution short-time Fourier transform (STFT) loss (with window size of 256, 512, 768, and 1024), and phase loss \cite{mp_senet} to ensure accurate waveform reconstruction, spectral consistency, and phase preservation. The time domain loss minimizes differences between the predicted and clean signals, while the multi-resolution STFT loss enforces consistency across multiple frequency resolutions, improving robustness against noise. The phase loss \cite{mp_senet} further refines phase estimation, reducing distortions and enhancing naturalness. The overall loss function is defined as:
\begin{equation}
\mathcal{L} = \lambda_1 \mathcal{L}_{\text{time}} + \lambda_2 \mathcal{L}_{\text{STFT}} + \lambda_3 \mathcal{L}_{\text{phase}},
\end{equation}
where $\lambda_1=0.5$, $\lambda_2=0.5$, and $\lambda_3=0.3$ are weighting factors balancing the contribution of each loss term.

\subsection{Feature Mapping and Sampling Frequency Independence}
The original SEMamba~\cite{chao2024investigation} used a masking-based approach, which applied predicted real-valued gain between 0 and 1 to the noisy magnitude spectrogram. While masking is easier to learn under the assumption of pure additive noise input, it falls short for distortions requiring content generation, such as bandwidth extension and packet loss, as the solution space extends beyond the masking-based approach. To better address reverberation and packet loss, we shifted from masking-based to mapping-based prediction. Specifically, the original sigmoid masking module was replaced with a Softplus-based magnitude estimator. USEMamba thus adopts a mapping-based strategy, directly learning the relationship between noisy and enhanced magnitude spectra to improve reconstruction quality.

To ensure generalization across different sampling rates with a single model, USEMamba integrates sampling frequency-independent (SFI) STFT and iSTFT \cite{zhang2023toward}. Unlike fixed STFT parameters, the model dynamically adjusts the FFT window and hop size length (fixed time duration) based on the input sampling rate, maintaining a consistent duration of feature frames across different sampling rates. 



\subsection{Efficient Scaling with Deeper Time-Frequency Mamba}
\begin{figure}[htb]
    \centering
    \centerline{\includegraphics[width=5.9cm]{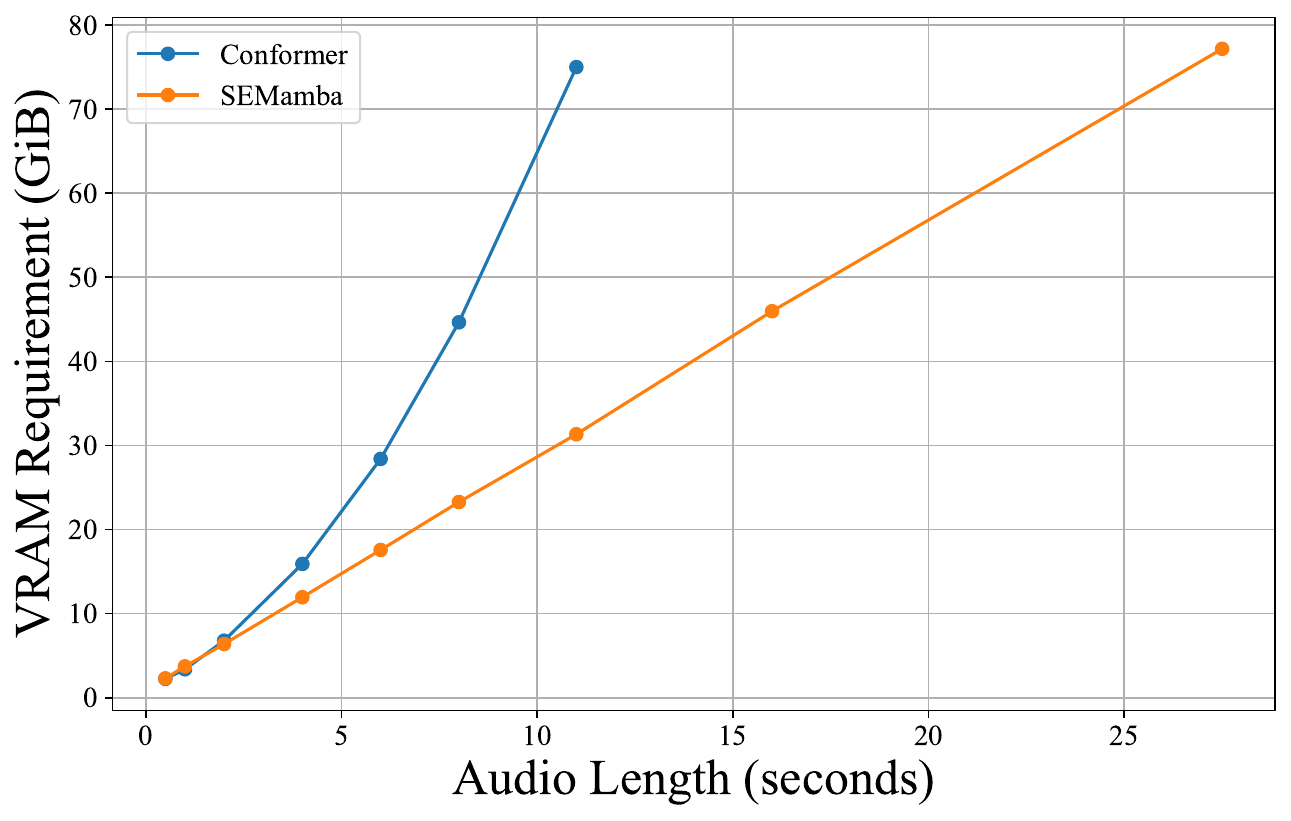}}
    \vspace{-0.6em}
    \caption{GPU VRAM requirement for 16kHz audio in training.} 
    \label{fig:audio_length}
\vspace{0mm}
\end{figure}

To improve performance while maintaining efficiency, we increase the number of Time-Frequency Mamba (TF-Mamba) layers to $N_1 = 10$ and the $N_2 = 4$. A key advantage of Mamba-based architectures over self-attention models is their memory efficiency, particularly in training, where storing gradient graphs notably increases VRAM consumption. Figure~\ref{fig:audio_length} illustrates the GPU VRAM requirement for 16 kHz audio during training. Transformer-based models, such as Conformer, experience notable increase in memory usage as input sequence length grows, limiting scalability. In contrast, Mamba-based architectures scale linearly, enabling deeper networks without excessive computational overhead.


\subsection{Generative USEMamba with Flow-Based Training}
To handle distortions which require audio content generation, such as bandwidth extension and packet loss \cite{li2024masksr}, we adopt a generative USEMamba variant using a flow matching-based model. A generative approach is particularly suitable for cases where missing speech components need to be inferred, e.g, for reconstructing realistic speech structures beyond the capabilities of purely regression-based methods.

In general, flow matching~\cite{lipman2023flow} aims to transform samples from a known, tractable distribution at $\tau=0$, e.g., standard normal distribution, into an typically intractable distribution of the target clean speech at $\tau=1$~\cite{liu2024generative, ku2025generative}. In practice, conditional flow matching is used for tractability~\cite{lipman2023flow}, conditioned on a target clean speech example $\mathbf{s}$ at $\tau=1$~\cite{liu2024generative, ku2025generative}. For current state $\mathbf{x}_\tau$, the conditional velocity field $v_\tau$ can then be computed as $v_\tau \left( \mathbf{x}_\tau | \mathbf{s}, \mathbf{y} \right) = \frac{\mathbf{s} - (1-\sigma_\text{min})\mathbf{x}_\tau}{1 - (1-\sigma_\text{min})\tau}$, where $\mathbf{s}$ are the clean speech coefficients in the compressed STFT domain, $\mathbf{y}$ are the observed degraded coefficients in the compressed STFT, and we use $\sigma_\text{min}=10^{-4}$. A neural model $u_\theta \left( \mathbf{x}_\tau, \mathbf{y}, \tau \right)$ with parameters $\theta$ is then trained to estimate the vector field by minimizing $\mathcal{L}_\text{CFM} = \mathbb{E}_{\left(\mathbf{s}, \mathbf{y} \right), \tau, \mathbf{x}_\tau} \left\lVert u_\theta \left( \mathbf{x}_\tau, \mathbf{y}, \tau \right) - v_\tau \left( \mathbf{x}_\tau | \mathbf{s}, \mathbf{y} \right) \right\rVert^2$, where $\left(\mathbf{s}, \mathbf{y} \right)$ are sampled from the training data, $\tau$ is sampled uniformly from $\left[0, 1\right ]$, and $\mathbf{x}_\tau$ is sampled from $\mathcal{N}\left(\tau\mathbf{s}, (1-t)^2\mathbf{I} \right)$.

The neural model $u_\theta$ is implemented using a USEMamba-based architecture by adding the magnitude and phase of $\mathbf{x}_\tau$ as additional channels to the feature encoder, analogously to the magnitude and phase of the noisy input $\mathbf{y}$. Time embedding is obtained using sinusoidal position encoding of $\tau$ with dimension set to 1024~\cite{liu2024generative, ku2024generative}. The embedding is used in USEMamba by adding adaptive normalization layers from diffusion transformers~\cite{peebles2023dit} before time- and frequency-Mamba as depicted in Fig.~\ref{fig:semamba_architecture}.

At inference time, an estimate of the clean coefficients $\hat{\mathbf{s}}$ in the compressed STFT domain for an observed degraded example $\mathbf{y}$ is generated by sampling $\mathbf{x}_0$ from standard normal distribution $\mathcal{N}\left(0, \mathbf{I}\right)$ and solving the ordinary differential equation determined by the estimated velocity field $\mathrm{d} \mathbf{x} = u_\theta(\mathbf{x}, \mathbf{y}, \tau) \mathrm{d}\tau$ to obtain $\hat{\mathbf{s}} = \mathbf{x}_1$. We use the Euler solver with a uniformly-spaced time grid and 20 inference steps.





\begin{table}[t]
    \centering
    \caption{Non-Blind Phase  Results of USEMamba variants. USEM stands for USEMamba.}
    \vspace{-0.6em}
    \resizebox{0.85\columnwidth}{!}{%
    \begin{tabular}{lccc}
        \toprule
        \textbf{Metric} & \textbf{USEM} & \textbf{USEM-Flow} & \textbf{Combined} \\
        \midrule
        PESQ $\uparrow$ & 2.79 & 1.54 & 2.77 \\
        ESTOI $\uparrow$ & 0.85 & 0.59 & 0.85 \\
        SDR $\uparrow$ & 13.11 & 4.49 & 13.08 \\
        MCD $\downarrow$ & 2.93 & 6.10 & 2.97 \\
        LSD $\downarrow$ & 2.94 & 3.91 & 2.94 \\
        DNSMOS $\uparrow$ & 3.01 & 2.50 & 3.00 \\
        NISQA $\uparrow$ & 3.21 & 2.82 & 3.19 \\
        UTMOS $\uparrow$ & 2.30 & 1.79 & 2.29 \\
        SBERT $\uparrow$ & 0.90 & 0.76 & 0.90 \\
        LPS $\uparrow$ & 0.85 & 0.61 & 0.85 \\
        SpkSim $\uparrow$ & 0.84 & 0.66 & 0.84 \\
        WAcc $\uparrow$ & 88.05 & 69.07 & 88.00 \\

        \bottomrule
    \end{tabular}
    }
    \label{tab:nonblind_results}
\end{table}

\begin{table*}[t]
    \centering
    \caption{Track 1 Blind Phase Results of the URGENT 2025 Challenge.}
    \resizebox{0.95\textwidth}{!}{%
    \begin{tabular}{lcccccccccccccc}
        \toprule
        \textbf{Team /}
        & \multicolumn{5}{c}{\textbf{Intrusive SE metrics}} 
        & \multicolumn{3}{c}{\textbf{Non-intrusive SE metrics}} 
        & \multicolumn{2}{c}{\textbf{Task-ind.}} 
        & \multicolumn{2}{c}{\textbf{Task-dep.}} 
        \\
        \cmidrule(lr){2-6}
        \cmidrule(lr){7-9}
        \cmidrule(lr){10-11}
        \cmidrule(lr){12-13}
        
        \textbf{Rank} & \textbf{PESQ} & \textbf{ESTOI} & \textbf{SDR} & \textbf{MCD} & \textbf{LSD} & \textbf{DNSMOS} & \textbf{NISQA} & \textbf{UTMOS} & \textbf{SBERT} & \textbf{LPS} & \textbf{SpkSim} & \textbf{WAcc} \\
        \midrule
        
        1th & 2.64 & 0.82 & 12.66 & 3.67 & 2.93 & 2.88 & 3.22 & 2.09 & 0.87 & 0.74 & 0.76 & 79.80 \\
        \textbf{Ours} & 2.48 & 0.81 & 11.69 & 3.64 & 2.98 & 2.83 & 2.92 & 2.03 & 0.85 & 0.72 & 0.74 & 78.35 \\
        3rd & 2.47 & 0.79 & 11.10 & 3.96 & 2.99 & 2.92 & 3.24 & 2.16 & 0.84 & 0.70 & 0.74 & 76.06 \\
        4th & 2.47 & 0.80 & 11.47 & 3.90 & 2.94 & 2.80 & 3.01 & 2.04 & 0.85 & 0.71 & 0.74 & 78.06 \\
        5th & 2.45 & 0.79 & 11.25 & 4.79 & 3.66 & 2.94 & 3.25 & 2.19 & 0.83 & 0.71 & 0.71 & 77.09 \\
        6th & 2.29 & 0.78 & 10.58 & 4.22 & 3.78 & 2.91 & 3.28 & 1.98 & 0.83 & 0.70 & 0.70 & 77.15 \\
        7th & 2.45 & 0.78 & 10.74 & 3.95 & 2.81 & 2.85 & 3.08 & 1.98 & 0.84 & 0.68 & 0.70 & 75.28 \\
        8th & 2.25 & 0.76 & 10.39 & 3.89 & 2.95 & 2.92 & 3.16 & 1.97 & 0.82 & 0.66 & 0.69 & 75.30 \\
        9th & 1.99 & 0.74 & 7.43 & 4.42 & 3.14 & 3.08 & 3.69 & 2.11 & 0.80 & 0.65 & 0.71 & 72.47 \\
        \bottomrule
    \end{tabular}
    }
    \label{tab:track1_results}
\end{table*}

\section{Experiments}

\subsection{Dataset}
The URGENT 2025 Challenge dataset consists of multi-condition speech recordings (across five languages: English, German, French, Spanish, and Chinese), noise samples, and room impulse responses (RIRs), as summarized in Table ~\ref{tab:dataset}. It considers seven types of distortions: additive noise, reverberation, clipping, bandwidth extension, codec artifacts, packet loss, and wind noise.  

Due to the short period of the Challenge and hardware constraints, we only used a \textbf{subset of the full training}
data (approximately 1.5 TB, as highlighted in bold in Table \ref{tab:dataset}). Specifically, regression USEMamba was trained exclusively on English speech data. 




\subsection{Evaluation Metrics}
We evaluated USEMamba using both intrusive and non-intrusive speech enhancement metrics. Perceptual Evaluation of Speech Quality (PESQ) assesses perceptual quality~\cite{Rix2001}, Extended Short-Time Objective Intelligibility (ESTOI) measures intelligibility~\cite{jensen2016stoi}, and signal-to-distortion ratio (SDR) evaluates distortion in the time-domain waveform~\cite{roux2019sisdr}. Mel Cepstral Distortion (MCD) and Log-Spectral Distance (LSD) capture spectral deviations between enhanced and reference speech. Non-intrusive metrics, including DNSMOS \cite{reddy2022dnsmos}, NISQA \cite{mittag2021nisqa}, and UTMOS \cite{saeki2022utmos}, provide perceptual speech quality estimates without requiring a clean reference. Additionally, results on downstream tasks were also evaluated. Task-independent metrics include SpeechBERTScore (SBERT) \cite{saeki2024speechbertscore}, which quantifies enhancement quality using a self-supervised model, and Levenshtein Phoneme Similarity (LPS)\cite{pirklbauer2023evaluation}, which measures the similarity between clean and enhanced phoneme sequences. Task-dependent metrics include speaker similarity (SpkSim), which evaluates the retention of speaker identity, and word accuracy (WACC), which reflects ASR performance.

\subsection{Non-Blind and  Blind Phases Results}
In Table \ref{tab:nonblind_results},  to examine the effectiveness of different model configurations, we evaluated the regression model (USEMamba), the generative model (USEMamba-Flow), and a combined approach where the
bandwidth extension and packet loss concealment parts of the regression model were replaced by the generative model. Despite being trained exclusively on English speech data, regression USEMamba demonstrated \textbf{strong generalization across multilingual test conditions}, achieving \textbf{2nd place in the non-blind test phase}. This highlights that the proposed USEMamba has language agnostic enhancement ability.


\begin{figure}[t]
 \subfloat[Distorted input]{\includegraphics[width=0.49\linewidth, height=2.0cm]{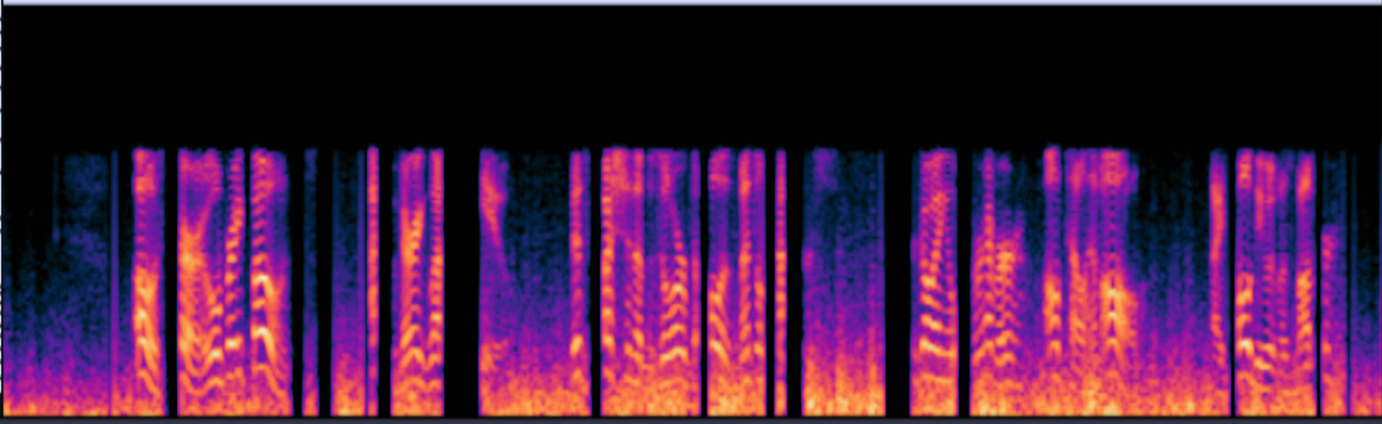}}
 \hfill
  \subfloat[USEMamba]{\includegraphics[width=0.49\linewidth, height=2.0cm]{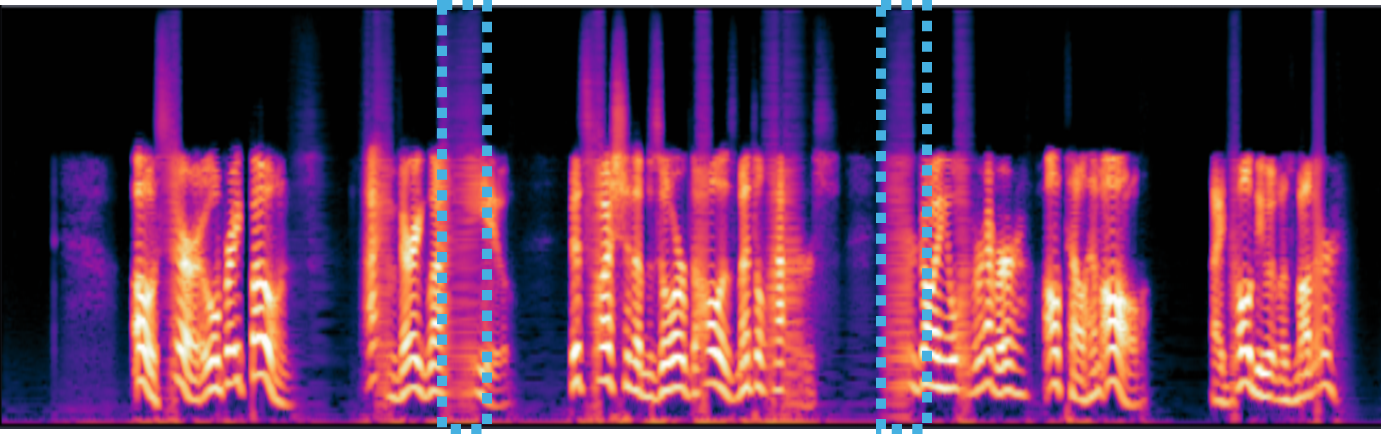}}
 \newline
  \subfloat[USEMamba-Flow]{\includegraphics[width=0.49\linewidth, height=2.0cm]{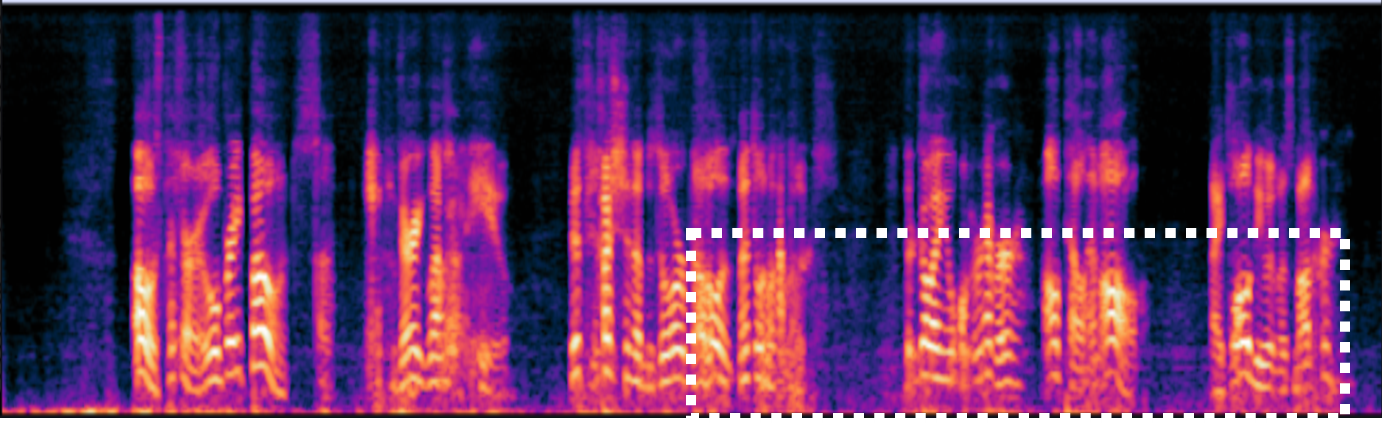}}
 \hfill
  \subfloat[Combined USEMamba with USEMamba-Flow]{\includegraphics[width=0.49\linewidth, height=2.0cm]{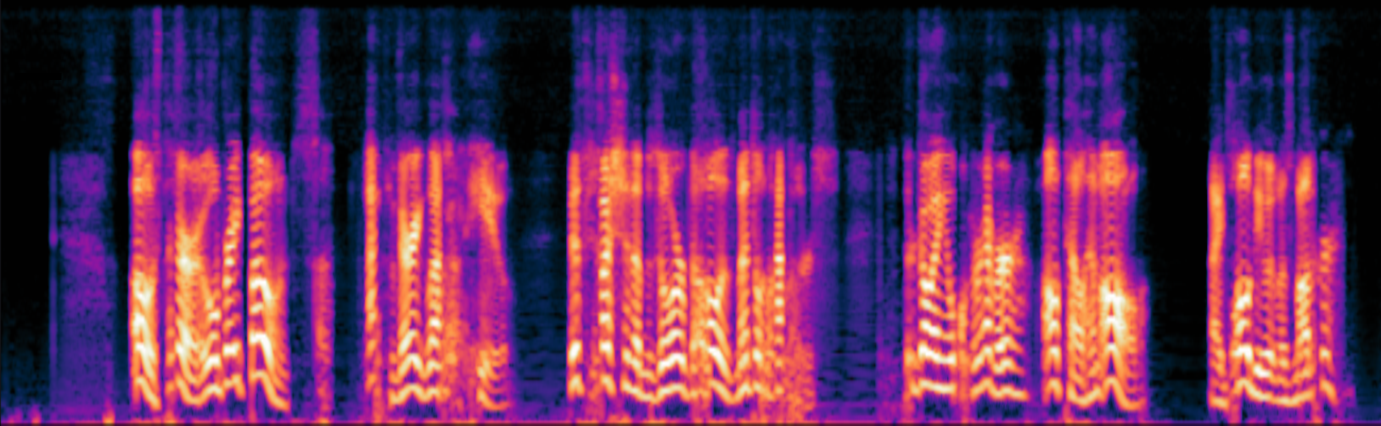}}
  \vspace{-0.6em}
  \caption{Spectrogram comparison of (a) distortion input (noisy, bandwidth limitation, and packet loss) and enhanced speech from (b) regression model (USEMamba), (c) generative model (USEMamba-Flow), and (d) proposed combined method.}
  \label{fig: compare}
\end{figure}

Although USEMamba with regression loss performed well in most signal recovery conditions, it generated over-smoothing results in cases requiring content generation, such as bandwidth extension and packet loss concealment, as indicated by the blue dashed rectangles in Figure \ref{fig: compare}.b. These distortions involve missing speech components that cannot be directly reconstructed through regression-based learning. The generative model (USEMamba-Flow), on the other hand, was able to mitigate such artifacts by inferring the missing content. However, its enhanced speech generally has low scores \cite{pirklbauer2023evaluation} evaluated by intrusive metrics such as PESQ, etc. In addition, we found that the outputs of USEMamba-Flow usually contain more residual noise (potentially due to insufficient training time given the tight schedule of this Challenge) compared to the regression baseline, as indicated by the white dashed rectangles in Figure \ref{fig: compare}.c. 

To keep the advantages of both regression and generative models, we propose a combined method based on a simple energy criterion of noisy input to determine the bandwidth extension and packet loss concealment regions $R$. Then the combined results are decided based on the following equation:
\[
Output_{t,f} = 
\begin{cases}
  \text{USEMamba}(Y)_{t,f} & \text{if $Y_{t,f}$  $\not\in R$ } \\
  \text{USEMamba-Flow}(Y)_{t,f} & \text{if $Y_{t,f}$  $\in R$ }
\end{cases}
\vspace{-0.3em}
\]
where $t$, and $f$ are the index of time and frequency index of the complex spectrogram, respectively. A spectrogram example is shown in Figure \ref{fig: compare}.d, which sounds more natural \footnote{In packet loss concealment cases, although the output sounds more natural, the phonemes are often incorrect when the packet loss duration is too long. Without phoneme conditioning (as in Voicebox \cite{le2024voicebox}) and considering multilingual settings, packet loss concealment remains a challenging task.} in the over-smoothed regions, while the objective scores in Table \ref{tab:nonblind_results} remain largely maintained. Finally, our model achieves \textbf{2nd place in the blind test phase} as shown in Table \ref{tab:track1_results}.

\vspace{-0.2em}
\section{Conclusion}
We presented Universal Speech Enhancement Mamba (USEMamba), a state-space model-based speech enhancement system designed to handle diverse distortions within the \textbf{Interspeech URGENT 2025 Challenge} framework. Despite being trained on only a subset of the full dataset (English data), USEMamba achieved \textbf{2nd place in the blind test phase}, and 5th place in the final ranking when the subjective SE metric was included, demonstrating strong generalization across various distortion types and languages. To mitigate the over-smoothing issue, we propose a combined approach that leverages both regression and generative models.

Future work will focus on enhancing the model’s performance under extreme distortions and achieving a better balance between regression and generative capabilities within a single USE model. Additionally, we will explore one of the key topics of this Challenge—the trade-off between the quality and quantity of training data.   

\vspace{-0.3em}
\section{Acknowledgment}
This work was supported by the NVIDIA Taiwan R\&D Center Sponsored Research program. The authors thank NVIDIA Corporation for their support.

\bibliographystyle{IEEEtran}
\bibliography{mybib}

\end{document}